\documentclass[aps,prl,showpacs,twocolumn]{revtex4}

\usepackage{graphicx}
\usepackage[ansinew]{inputenc}
\usepackage{color}
\usepackage{float}
\definecolor{Blue}{rgb}{0.3,0.3,0.9}
\newcommand{\Figref}[1]{Fig.~\ref{#1}}
\newcommand{\Eqref}[1]{Eq.~(\ref{#1})}

\begin{document}

\title{Model study of the electron-phonon coupling in graphene; relative importance of intraband and interband scattering}

\author{H. Toren}
\affiliation{Department of Physics, University of Gothenburg, Sweden}

\author{L. Samuelsson}
\affiliation{Department of Physics, University of Gothenburg, Sweden}

 \author{B. Hellsing}
\affiliation{Department of Physics, University of Gothenburg, Sweden}

\begin{abstract}
The aim of this model study of the electron-phonon coupling in graphene was to find out about the relative importance of the inter- and intraband scattering and which phonon modes are the most active. This was achieved by analyzing the electron-phonon matrix element of the carbon dimer in the unit cell. We found that for the intra molecular orbital matrix elements the longitudinal optical phonon mode is the active phonon mode. The matrix element corresponding to $\sigma \rightarrow \sigma$ is greater than the matrix element for $\pi \rightarrow \pi$ . The inter molecular orbital scattering $\pi \rightarrow \sigma$ is driven by the out-of-plane acoustic phonon mode, while the out-of-plane optical mode does not contribute for symmetry reasons. We found the unexpected result that the magnitude of matrix element of the inter molecular orbital scattering $\pi \rightarrow \sigma$ exceeds the intra molecular orbital scattering $\pi \rightarrow \pi$. These results indicate that the in general not considered inter-band scattering has to be taken into account when analyzing e.g. photo-hole lifetimes and the electron-phonon coupling constant $\lambda$ from photoemission data of graphene.
\end{abstract}



\maketitle

\section{Introduction}

Several experimental and theoretical studies of graphene have been presented during the last decade.\cite{Neto09,Sarma11} These investigations have revealed remarkable mechanical \cite{Lee08}, electronic \cite{Morozov08}, optical \cite{Bonaccorso10} and thermal \cite{Balandin08} properties. However, unstrained pristine graphene is not considered to be a good superconductor due to a weak electron-phonon coupling (EPC). The presented simple model study confirms the weak EPC in the $\pi$ bands while the coupling should be considerably stronger in the occupied part of the $\sigma$ band.

Our investigation is inspired by the recently reported experimental work by Mazzola et al.. \cite{Mazzola_13} The evidence of a strong EPC in the $\sigma$-band was revealed in angle-resolved photoemission spectroscopy (ARPES) measurements with a substantial lifetime broadening and a pronounced characteristic kink in the band dispersion. The EPC constant $\lambda$ was determined to be approximately 1 near the top of the $\sigma$ band. \cite{Mazzola_13} This is about a factor 3-4 larger than what have been found in the $\pi$ band.

The aim of this tight-binding model study has been to sort out, (1) the relative importance of the interband and intraband scattering and, (2) which phonon mode contributes the most in theses scattering processes. The study is focused on calculations and analysis of the electron-phonon (e-ph) matrix element $M_{e-ph}$, which is the core of the EPC. Approximately we have that $\lambda \sim |M_{e-ph}|/\Omega_{0}$, where $\Omega_{0}$ is the dominant phonon mode frequency.

\section{Theory}

The EPC constant $\lambda$ can be examined in ARPES measurements. The created photo-hole yields a non-equilibrium which drives electron scattering in order to fill the hole. According to the Heisenberg uncertainty relation the energy width of the detected photo-electron kinetic energy - the lifetime broadening - is inversely proportional to the hole lifetime. At low temperatures, electron scattering is assisted by phonon emission. Theoretical modeling in order to calculate the phonon induced lifetime broadening enables estimation of the strength of the EPC determined by $\lambda$. Experiments indicate that the high energy optical phonon modes are the most active modes in the EPC in graphene. \cite{Mazzola_13,Haberer_K_point_2013} 

The strength of the EPC is determined by the e-ph matrix element 
\begin{equation}
M^\nu_{e-ph}=\left<\Phi_{n'\mathbf{k}'}(\mathbf{r})|\delta V^{\nu\mathbf{q}}(\mathbf{r}) |\Phi_{n\mathbf{k}}(\mathbf{r})\right> \ ,
\label{eq:ep_mat}
\end{equation}
where $\Phi_{n\mathbf{k}}$ and $\Phi_{n'\mathbf{k'}}$ are initial and final state wave functions of bands $n$ and $n'$, respectively. $\delta V^{\nu \mathbf{q}}$ is the phonon-induced deformation potential where $\nu$ and $\mathbf{q}$ labels phonon mode and momentum, respectively.

\subsection{Tight-binding model}
In the tight-binding approximation the valence wave functions of graphene are formed by superposition of the atomic $2s$ and $2p$ orbitals. The $\sigma$ band ($n = \sigma$) states are built by $2s$, $2p_x$ and $2p_y$ orbitals and the $\pi$ band ($n = \pi$) by the $2p_z$ orbital. The tight-binding wave functions is give by
\begin{equation}
\Phi_{n\mathbf{k}}(\mathbf{r})=\sum_{u s j}c_{nsj}(\mathbf{k})\phi_{nsj}(\mathbf{r}-\mathbf{R}_{us})e^{i\mathbf{k}\mathbf{R}_{us}} \ ,
\label{eq:wave}
\end{equation}
%
%
\begin{figure}
\includegraphics[angle=0,origin=c,width=80 mm]{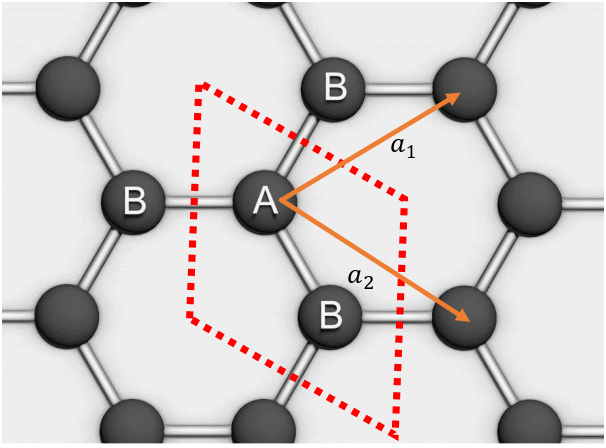}
\caption{Schematic drawing of the graphene unit cell with the two carbon atoms denoted by $A$ and $B$ and enclosed by the dashed line. The two lattice vectors $a_{1}$ och $a_{2}$ are shown.}
\label{unit_cell}
\end{figure}

where $u$ labels the unit cells hosting the two different carbon atom sites, $s=A,B$ and $\mathbf{R}_{us}$ denote the equilibrium atom positions. The index $j$ labels the atomic site orbital. The orbital $\phi_{nsj}$ then refers to the orbital $j$ of atom type $s$ of the band $n$. In the case of the occupied bonding $\pi$ band the site orbital is the same at all sites $\phi_{2p_z}$ while for the $\sigma$ band we have the three differently oriented $sp^{2}$ orbitals.
 
The deformation potential is given by
\begin{equation}
\delta V^{\nu \mathbf{q}}(\mathbf{r})= - \sum_{us}\mathbf{u}^{\nu\mathbf{q}}_{us}\cdot\mathbf{\nabla}v(\mathbf{r}-\mathbf{R}_{us}) \ ,
\label{eq:dv}
\end{equation}
where $\mathbf{q}=\mathbf{k}-\mathbf{k'}$ and 
\begin{equation}
\mathbf{u}^{\nu\mathbf{q}}_{us}= \frac{1}{2}A_{\nu\mathbf{q}}\mathbf{e}^{\nu}_{s}(\mathbf{q}) e^{i\mathbf{q}\cdot\mathbf{R}_{us}} \ ,
\label{eq:uvec}
\end{equation}
where \( A_{\nu\mathbf{q}}=\sqrt{\hbar/Nm\omega_{\nu}(\mathbf{q})} \) is the amplitude of phonon mode $\nu$, with $N$ the number of unit cells, $m$ atomic mass and $\mathbf{e}^{\nu\mathbf{q}}_{s}$ the phonon polarization vector. 
In the case when the temperature is low enough that only phonon emission has to be considered, which is the case in the experiment by Mazzola et al. \cite{Mazzola_13}, we have
\begin{equation}
M^{\nu,nn'}_{e-ph}= - \frac{1}{2}A_{\nu\mathbf{q}}D^{\nu}_{nn'}(\mathbf{k},\mathbf{k'}) \delta_{\mathbf{k}-\mathbf{k}', \ \mathbf{q}} \ ,
\label{eq:M2}
\end{equation}
where 
\begin{eqnarray}
D^{\nu}_{nn'}(\mathbf{k},\mathbf{k}') &=& \sum_{uu'u''}\sum_{ss's''} \sum_{jj'} c_{n's'j'}^*(\mathbf{k}')c_{nsj\mathbf{k}}(\mathbf{k}) \times \nonumber \\ 
&& \mathbf{m}_{s''}^{nj,n'j'}(\mathbf{R}_{u's'},\mathbf{R}_{us})\cdot \mathbf{e}^{\nu}_{s''}(\mathbf{q})  \times \nonumber \\ 
&& e^{-i\mathbf{k'}\cdot(\mathbf{R}_{u's'}-\mathbf{R}_{u''s''})}e^{i\mathbf{k}\cdot(\mathbf{R}_{us}-\mathbf{R}_{u''s''})} ,
\label{eq:D}
\end{eqnarray}
where the site e-ph matrix element is given by the local site orbitals and the site dependent deformation potential
\begin{eqnarray}
&&\mathbf{m}_{s''}^{nj,n'j'}(\mathbf{R}_{u's'},\mathbf{R}_{us}) = \nonumber \\
&&\int \phi_{n's'j'}^*(\mathbf{r}-\mathbf{R}_{u's'}) \times \nonumber \\
&&\mathbf{\nabla}v(\mathbf{r}-\mathbf{R}_{u''s''})\phi_{nsj}(\mathbf{r}-\mathbf{R}_{us})d\mathbf{r} \ ,
\label{eq:m}
\end{eqnarray}
The vibrational amplitude of the individual carbon atoms is at least an order of magnitude less than the lattice parameter. Thus we expect that the greatest contribution to the three center integral in \Eqref{eq:m} is expected for the case when $\mathbf{R}_{u's'}$ and $\mathbf{R}_{us}$ correspond to pairwise nearest neighboring $A$ and $B$ sites and $\mathbf{R}_{u''s''}$=$\mathbf{R}_{u's'}$ or $\mathbf{R}_{u''s''}$=$\mathbf{R}_{u's'}$. The three center integral is then reduced to a two center integral. However, it has been pointed out that for a realistic deformation potential, the convergence with respect to order of neighboring sites is surprisingly slow.\cite{Jiang_04}. 

For our purpose to qualitatively understand the relative importance of the intraband and interband scattering and the most important phonon mode in the two cases, we simplify furthermore and assume that $u=u'$ referring to the unit cells of the atomic site wave functions. Thus we consider only intersite scattering within the unit cell. 

We will take the matrix element of $D$ as a measure of the strength of the EPC. According to \Eqref{eq:M2} we have that $|M^{\nu,nn'}_{e-ph}|^{2} \sim |D^{\nu}_{nn'}|^2/\Omega_0$, where $\Omega_{0}$ is a presumably dominant phonon mode frequency. This connects to the ECP strength measured by $\lambda$, which is determined by the reciprocal frequency moment of the Eliashberg function which leads to the additional frequency factor in the denominator, $\lambda \sim (|D^{\nu}_{nn'}|/\Omega_{0})^{2}$. 

To simplify the notation to express $D^{\nu}_{nn'}$ in \Eqref{eq:D} we introduce the two molecular orbitals corresponding to the binding $\sigma$ and $\pi$ orbitals. We refer to a coordinate system with the $z$ axis oriented in the direction of the $AB$ dimer of the unit cell and the $x$ axis in the direction normal to the graphene layer. With $B$ atom in the origin and $A$ atom in the position $z$ = $R_A$ the orbitals are written 
\begin{eqnarray}
\Psi_\sigma(\mathbf{r})&=&\left|sp^2_z\right>_A+\left|sp^2_z\right>_B \nonumber \\
&=&\left|2s\right>_A-\left|2p_z\right>_A+\left|2s\right>_B+\left|2p_z\right>_B \ ,
\label{eq:sigma}
\end{eqnarray}
and 
\begin{equation}
\Psi_\pi(\mathbf{r})=\left|2p_x\right>_A+\left|2p_x\right>_B
\label{eq:pi}
\end{equation}
where the low index $A$ and $B$ refers to the location of the atomic wave functions in the unit cell. The normalized molecular wave functions are then given by
\begin{equation}
\Phi_n(\mathbf{r})=\frac{\Psi_n(\mathbf{r})}{\left<\Psi_n|\Psi_n\right>}.
\end{equation}
For the case $u=u'=u''$ and omitting phase factors, which will be of types $e^{i\mathbf{k'}\cdot\mathbf{R}_A}$ and $e^{i\mathbf{k}\cdot\mathbf{R}_A}$ in \Eqref{eq:D}, the $D$ matrix takes the form
\begin{eqnarray}
\tilde{D}^{\nu}_{nn'} = \left<\Phi_n|\nabla v_\nu|\Phi_{n'}\right> .
\label{eq:Dm}
\end{eqnarray}
The diagonal elements of the $\tilde{D}$ matrix corresponds to the intraband scattering $\tilde{D}^{\nu}_{\sigma \sigma}$ and $\tilde{D}^{\nu}_{\pi \pi}$, while the off-diagonal elements $\tilde{D}^{\nu}_{\sigma \pi}$ and $\tilde{D}^{\nu}_{\pi \sigma}$ correspond to interband scattering. 

The deformation potential $\nabla v_\nu$ referring to the three optical phonon modes (ZO, TO and LO) and the three acoustical modes (ZA, TA and LA) are generated by taking the gradient of the one-electron atomic potential $V$ at the $A$ and $B$ sites as follows

\begin{eqnarray}
 \nabla v_{ZO} &=& -\frac{\partial V}{\partial x}\Big|_A + \frac{\partial V}{\partial x}\Big|_B \ , \ 
 \nabla v_{ZA} = \frac{\partial V}{\partial x}\Big|_A + \frac{\partial V}{\partial x}\Big|_B \nonumber \\ 
 \nabla v_{TO} &=& -\frac{\partial V}{\partial y}\Big|_A + \frac{\partial V}{\partial y}\Big|_B \ , \ \nabla v_{TA} = \frac{\partial V}{\partial y}\Big|_A + \frac{\partial V}{\partial y}\Big|_B \nonumber 
 \\ 
 \nabla v_{LO} &=& -\frac{\partial V}{\partial z}\Big|_A + \frac{\partial V}{\partial z}\Big|_B \ , \ \nabla v_{LA} = \frac{\partial V}{\partial z}\Big|_A + \frac{\partial V}{\partial z}\Big|_B. \nonumber 
\label{ekv:deriv}
\end{eqnarray}

\section{Calculation}

The radial part of the $2s$ and $2p$ orbitals are chosen to be of Slater type \cite{Slater_orb}
\begin{equation}
R(r) \sim re^{-\frac{Z^{*}r}{2}},
\end{equation}
where the effective nuclear charge is $Z^{*}$=3.25 a.u. \cite{slater}. 

In order to minimize the number of parameters while hopefully retain the essential physics we assume that the on-site effective one-electron potential $V$ is spherically symmetric and of the form of a screened coulomb potential 
\begin{equation}
V(\mathbf{r})=V(r)=-V_0 e^{-\alpha r}/r \ , 
\end{equation}
where  $V_0$ is a positive constant and $\alpha$ is an inverse screening length. We then have an analytical expression of the deformation potential $\nabla v_\nu$ determined by \( \partial V/\partial x_{i}=x_{i}(\alpha+1/r)V(r) \), where $(x_1,x_2,x_3)=(x,y,z)$.

\section{Results}
We have performed numerical calculations of the matrix elements $D_{n,n'}^\nu = \left<\Psi_n|\nabla v_\nu|\Psi_{n'}\right>$. The contributions from the optical phonon mode LO dominates completely the intraband scattering, while the interband scattering is assisted by the acoustic mode ZA. For this reason the corresponding matrix elements are analyzed in more detailed. The screening constant is set according to  $\alpha=0.253\,\mathrm{a.u}^{-1}$. This value corresponds to the inverse of the most probable radius of the 2s electron in the free carbon atom. The distance between the $A$ and $B$ atoms in the unit cell is $R_A=2.683\,\mathrm{a.u}$ and $V_0$ = 1 a.u.. The numerical value of $V_0$ is of no importance as the e-ph matrix elements all scales with $V_0$ and we are only interested in the relative magnitude of the matrix elements.

\subsection{Intraband scattering}
In this section we analyze the intraband scattering by numerically calculating the diagonal matrix elements $\left<\Phi_n|\nabla v_\nu|\Phi_n\right>$, where $n=\{\sigma,\pi\}$.

\subsubsection{$\sigma\rightarrow\sigma$}
The contributions from the different phonon modes to this intraband scattering is shown in table \ref{table:sigma}. Evidently the dominant contribution is due to the LO phonon mode. The other phonon modes gives essentially zero contributions (numerically $<$ 10$^{-15}$). In table \ref{tab:sigsig} we show the LO driven scattering resolved in terms of the scattering between the atomic 2s and 2p$_z$ orbitals. 

\begin{table}
\caption{Numerical results for the contribution from all the phonon modes to the intraband scattering ($\sigma\to\sigma$)} 
\centering 
\begin{tabular}{c | c } 
 \hline\hline
 \\ 
\ \large{$\nu$} &  \ \ \large{$\left<\Phi_\sigma|\nabla v_{\nu}|\Phi_\sigma\right>$}  \\ 
\\
\hline 
 ZO &  0 \\ 
 TO &  0 \\
 LO &  0.71 \\
 ZA &  0 \\
 TA &  0 \\
 LA &  0 \\ [1ex]
\hline 
\end{tabular}
\label{table:sigma} 
\end{table}
%
The results show that all inter atomic orbital scattering is of the same order of magnitude. However the scattering between $\left|2p_{z}\right>_A\rightleftharpoons\left|2p_{z}\right>_B$, $\left|2s\right>_A\rightleftharpoons\left|2p_{z}\right>_A$ and $\left|2s\right>_B\rightleftharpoons\left|2p_{z}\right>_B$ give the largest contributions to the matrix element.
\begin{table}
\caption{The atomic orbital contributions to the intraband scattering matrix element ($\sigma\rightarrow\sigma$) for the LO phonon mode.}
\centering
\label{tab:sigsig}
\begin{tabular}{ c }
 \hline \hline
 \\
\large{$\left<\Phi_\sigma|\nabla v_{LO}|\Phi_\sigma\right>\times10^2$} \\
  \\
  \begin{tabular}{ c | c c c c | c}  
    \hline
    Orbital & $\left|2s\right>_A$ & $\left|2s\right>_B$ & $\left|2p_{z}\right>_A$ & $\left|2p_{z}\right>_B$ & Sum \\
    \hline
    $\left|2s\right>_A$         &     1.08  &  1.44  &  7.72  &  3.35   &   13.59   \\
    $\left|2s\right>_B$         &     1.44  &  1.08  &  3.35  &  7.72   &   13.59   \\
    $\left|2p_{z}\right>_A$     &     7.72  &  3.35  &  3.03  &  8.05   &   22.14   \\
    $\left|2p_{z}\right>_B$     &     3.35  &  7.72  &  8.05  &  3.03   &   22.14   \\
   \hline
    \multicolumn{5}{c}{} & 71.46   \\
   \hline
  \end{tabular} 
  \end{tabular}
\end{table}
%
%
\subsubsection{$\pi\rightarrow\pi$}
The phonon mode contributions to the intraband scattering $\pi\to\pi$ is shown in table \ref{table:pi}. Just as in the case of the $\sigma\to\sigma$, the LO dominates completely. The corresponding inter atomic orbital scattering contribution is shown in table \ref{tab:pipi} and the order of magnitude is similar for all the inter orbital scattering. 
\begin{table}
\caption{Numerical results for the contribution from all the phonon modes to the intraband scattering ($\pi\to\pi$)} 
\centering 
\begin{tabular}{c | c } 
 \hline\hline
 \\
\ $\nu$ &  \ \ $\left<\Phi_\pi|\nabla v_{\nu}|\Phi_\pi\right>$ \\
\\
\hline 
ZO &  0 \\
TO &  0 \\
LO &  0.12 \\
ZA &  0 \\
TA &  0 \\
LA &  0 \\
\hline
\end{tabular}
\label{table:pi} 
\end{table}

It is not surprising that only the LO phonon mode is in operation for the intraband scattering as the products $\Phi_\sigma \otimes\nabla v_{LO}\otimes\Phi_\sigma$ and $\Phi_\pi\otimes\nabla v_{LO}\otimes\Phi_\pi$ are totally symmetric with respect to reflection in ($x,y$), ($x,z$) and ($y,z$) planes, while the other phonon modes fail this requirement. However, the interesting point is the difference in magnitude of the matrix elements, $|\left<\Phi_\sigma|\nabla v_LO|\Phi_\sigma\right>|^2$ $\approx$ 0.5 and $|\left<\Phi_\pi|\nabla v_LO|\Phi_\pi\right>|^2$ $\approx$ 0.01, which indicates an order of magnitude stronger EPC strength for the $\sigma$ intraband scattering compared to the $\pi$ intraband scattering. 

\begin{table}[h]
\caption{The atomic orbital contributions to the intraband scattering matrix element ($\pi\rightarrow\pi$) driven by the LO phonon mode.}
\centering
\label{tab:pipi}
\begin{tabular}{ c }
 \hline \hline
 \\
\large{$\left<\Phi_\pi|\nabla v_{LO}|\Phi_\pi\right>\times10^2$} \\
  \\
  \begin{tabular}{ c | c c | c}  
    \hline
    Orbital & $\left|2p_{x}\right>_A$ & $\left|2p_{x}\right>_B$ & Sum \\
    \hline
    $\left|2p_{x}\right>_A$     &    3.65  &  2.44  & 6.09  \\
    $\left|2p_{x}\right>_B$     &    2.44  &  3.65  & 6.09  \\
   \hline
    \multicolumn{3}{c}{} & 12.18   \\
  \end{tabular} 
  \\ \\
  \end{tabular}
\end{table}
%
%
\subsection{Interband scattering}
The contributions from different phonon modes to the interband scattering is shown in Table \ref{tab:inter}.

\begin{table}
\caption{Numerical results for the contribution from all the phonon modes to the interband scattering ($\pi\to\sigma$)} 
\centering 
\label{tab:inter}
\begin{tabular}{c | c } 
 \hline\hline
 \\ 
\ $\nu$ &  \ \ $\left<\Phi_\sigma|\nabla v_{\nu}|\Phi_\pi\right>$ \\
\\
\hline 
ZO &  0 \\
TO &  0 \\
LO &  0 \\
ZA &  0.38 \\
TA &  0 \\
LA &  0 \\
\hline
\end{tabular}
\end{table}
The overall dominant contribution is due to the out-of-plane acoustic phonon mode ZA. The contributions to the e-ph matrix element from scattering between the different atomic orbitals is revealed in Table \ref{tab:pisig}.  

\begin{table}
\caption{The atomic orbital contributions to the interband scattering matrix element ($\sigma\rightarrow\pi$) for the ZA phonon mode.}
\centering
\label{tab:pisig}
\begin{tabular}{ c }
 \hline \hline
 \\
\large{$\left<\Phi_\sigma|\nabla v_{LO}|\Phi_\pi\right>\times10^2$} \\
  \\
  \begin{tabular}{ c | c c c c | c}  
    \hline
    Orbital & $\left|2s\right>_A$ & $\left|2s\right>_B$ & $\left|2p_{z}\right>_A$ & $\left|2p_{z}\right>_B$ & Sum \\
    \hline
    $\left|2p_{x}\right>_A$     &   10.48  &  2.71  &  0.76  &  5.00 &  18.94   \\
    $\left|2p_{x}\right>_B$     &    2.71  & 10.48  &  5.00  &  0.76 &  18.94   \\
    \hline
    \multicolumn{5}{c}{} & 37.88  \\
\hline
  \end{tabular}
\end{tabular}
\end{table}

The largest contributions is accounted for by the scattering $\left|2p_x\right>_A\to\left|2s\right>_A$ och $\left|2p_x\right>_B\to\left|2s\right>_B$.

We note that for the interband scattering the $ZA$ phonon mode is the only mode that contributes to e-ph coupling, again the reason is that the $ZA$ mode is the only mode that gives a totally symmetric matrix element, $\Phi_\sigma\otimes\nabla v_{ZA}\otimes\Phi_\pi$. Furthermore, $|\left<\Phi_\sigma|\nabla v_{ZA}|\Phi_\pi\right>|^2$ $\approx$ 0.14, which indicates that the EPC strength of the interband scattering is of the same order of magnitude as the intraband scattering processes. 
In more details, the numerical calculation of the matrix elements indicate that the interband scattering is even stronger than the intraband $\pi$ scattering. 

\begin{figure}
\centering
\includegraphics[angle=0,origin=c,width=80 mm]{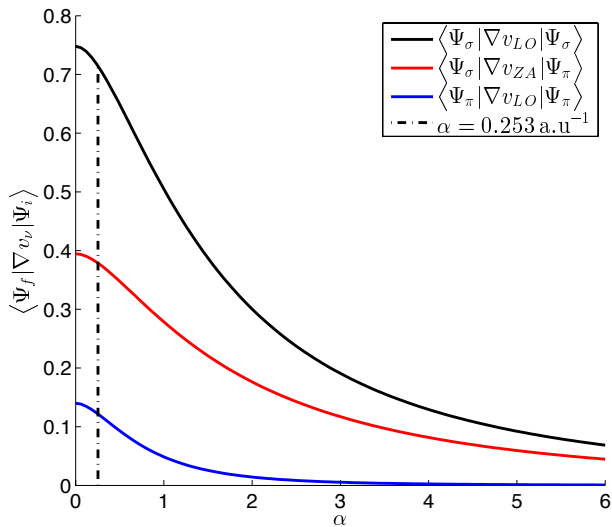}
\caption{The three dominant matrix elements as function of the $\alpha$. The dashed line corresponds to a presumably reasonable value for the screening parameter, $\alpha=0.253\,\text{a.u}^{-1}$.  }
\label{fig:varalpha}
\end{figure}

In order to find out about the sensitivity of these results referring to the single screening parameter ($\alpha$) of the one-electron model potential we performed calculations of the three matrix elements as a function of $\alpha$. The results shown in \Figref{fig:varalpha} show that the ordering of the magnitude of the matrix elements is kept in a wide range of values of the $\alpha$ parameter. The steady decrease of the matrix elements with $\alpha$ reflects the reduced screening length of the on site potential which will reduce the effectively integrated volumes in the calculations of the matrix elements 
\begin{figure}
\centering
\includegraphics[angle=0,origin=c,width=80 mm]{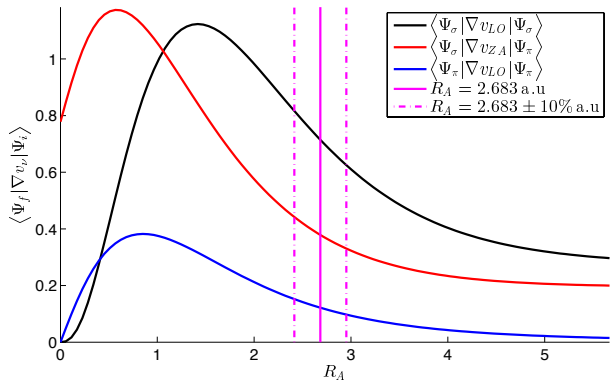}
\caption{The intra- and interband matrix elements as a function of $R_{A}$. The vertical solid line represent the inter atomic distance in graphene, $R_A$ =2.683 a.u.. The dashed dotted vertical lines show the change of $R_{A}$ in range corresponding $\pm10\%$.}
\label{fig:varr}
\end{figure}

Finally we investigated the sensitivity of the distance between the $A$ and $B$ site, $R_{A}$. The aim is to get some hint about the strain dependence of the EPC strength. The results shown in \Figref{fig:varr} show that changes of $R_{A}$ within 10 $\%$ (the range between the vertical dotted dashed lines) do not change the ordering of the three matrix elements. Furthermore the matrix elements decrease monotonically in the range due to the reduced overlap between the $A$ and $B$ atom.

Taking a step further with this tight binding approach and consider nearest neighbor hopping when calculating the band state wave functions in \Eqref{eq:wave}, $\Phi_{\sigma\mathbf{k}}(\mathbf{r})$ and $\Phi_{\pi\mathbf{k}}(\mathbf{r})$ amounts to diagonalizing a 6x6 and 2x2 matrix, respectively. Then for $\sigma$ intra band scattering and the interband scattering several additional atomic orbital matrix elements will come in to play. However, still the electron-phonon matrix elements we have analyzed within the unit cell will be the dominant type of matrix elements, simply due to large on-site wave function overlap. 
\section{Summary and discussions}
In this work the investigation is focused on the dominant electron-phonon matrix elements in graphene. In particular we are interested in getting a hint about the relative importance of the intraband and interband electron scattering and to what extent the different vibrational modes is in operation driving the scattering. For this reason we simplify drastically the tight binding calculation and consider only the $\sigma$ and $\pi$ bonding orbitals of the carbon dimer in the unit cell. \\ 
In addition to the well known fact that the optical LO mode dominates the intraband scattering $\sigma\rightarrow\sigma$ and $\pi\rightarrow\pi$, we find that the interband scattering $\pi \rightarrow \sigma$ is expected to be surprisingly strong. For symmetry reasons the only phonon mode driving the interband scattering is the acoustic ZA mode. For a realistic model deformation potential we find that for the carbon dimer the inter molecular orbital scattering is stronger than the intra molecular orbital scattering $\pi \rightarrow \pi$ while weaker than the intra molecular orbital scattering $\sigma \rightarrow \sigma$. In addition, these results persist for any reasonable in-plane strain in the graphene lattice. \\
The implication of these results is that in the energy region where the $\sigma$ and the $\pi$ electronic bands overlap we can expect that the interband scattering will contribute significantly, in addition to the intraband scattering, to the electron-phonon coupling constant $\lambda$. This in turn might be part of the reason for the experimentally large large $\lambda$ value near the top of the $\sigma$ band. Near the Fermi level there is no overlap between the $\pi$ band and the $\sigma$ band and thus no contribution to $\lambda$ from interband scattering.
%

\end{document}